\documentclass[
    ,final            
  ]
  {aipproc}

\layoutstyle{8x11double}

\setcitestyle{numbers}
\usepackage[dvipsnames]{color}
\usepackage{amsmath}
\usepackage{amssymb}

\begin{document}

\newcommand{\Imag}{\mathop{\mathrm{Im}}}

\title{A Strongly Interacting Electroweak Symmetry Breaking Sector with a Higgs-like light scalar}

\classification{11.15.Ex,11.30.Rd,11.80.Et,11.80.Gw,12.60.Cn,12.60.Fr}
\keywords      {Electroweak sector, Higgs boson, Strong Interactions, Unitarization, Equivalence Theorem}

\author{Rafael L. Delgado (speaker)}{
  address={On leave at University of Southampton, Southampton High Energy Physics theory group (SHEP), School of Physics and Astronomy, SO17 1BJ Highfield Southampton, UK. Permanent address Universidad Complutense de Madrid, Depto. de F\'{\i}sica Te\'orica I, Fac. CC. F\'{\i}sicas, Parque de las Ciencias 1, 28040 Madrid, Spain.}
}

\author{\\ Antonio Dobado}{
  address={Universidad Complutense de Madrid, Depto. de F\'{\i}sica Te\'orica I, Fac. CC. F\'{\i}sicas, Parque de las Ciencias 1, 28040 Madrid, Spain.}
}


\author{Felipe J. Llanes--Estrada}{
  address={Universidad Complutense de Madrid, Depto. de F\'{\i}sica Te\'orica I, Fac. CC. F\'{\i}sicas, Parque de las Ciencias 1, 28040 Madrid, Spain.}
}


\begin{abstract}
The apparent finding of a 125\,GeV light Higgs boson would close the minimal Standard Model (SM), that is weakly interacting. This is an exceptional feature not generally true if new physics exists beyond the mass gap found at the LHC up to 700 GeV.

Any such new physics would induce departures from the SM in the low-energy dynamics for the minimal electroweak symmetry-breaking sector (EWSBS), with three Goldstone bosons (related to longitudinal W and Z bosons) and one light Higgs-like scalar.

With no new particle content, for most of the parameter space, the scattering is actually strongly interacting (with the SM a remarkable exception). We therefore explore various unitarization methods, that have already be applied to the tree-level W$_L$ W$_L$ amplitude; we find and study a natural second sigma-like scalar pole there. Of note is its appearance due to either elastic or coupled-channel dynamics, especially since the later is largely unconstrained by current LHC data and could be large.

\end{abstract}

\maketitle

\section{Introduction}
\begin{figure}[!h]
\null\hfill\includegraphics[width=.33\textwidth]{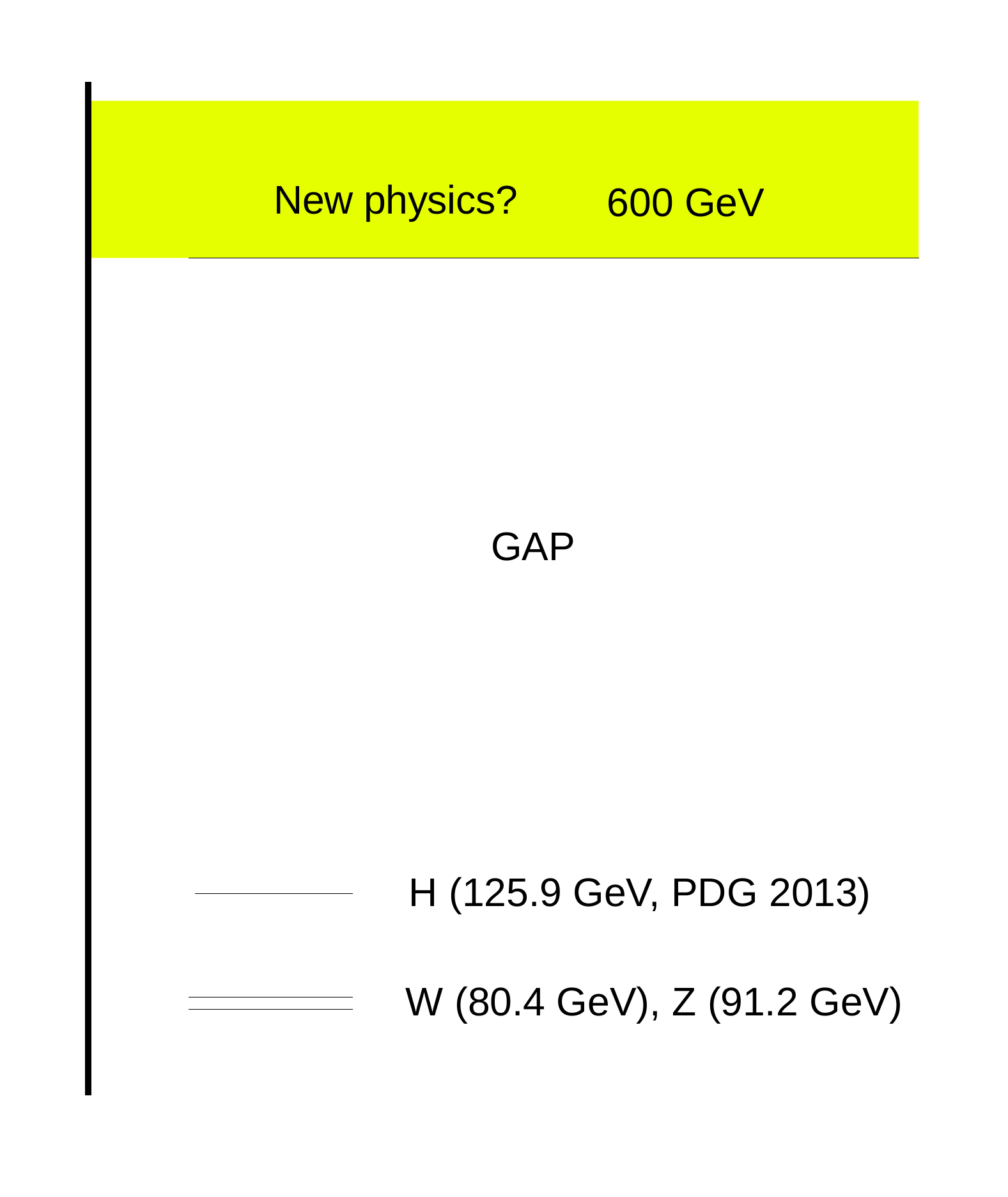}\hfill\null\\
\caption{EWSBS of the Standard Model after the LHC run I: there are four \emph{low-energy} bosons (3 massive bosons and the light Higgs-like scalar), and any new physics is beyond a mass gap.
}\label{gap}
\end{figure}

According to ATLAS~\cite{ATLAS} and CMS~\cite{CMS}, at LHC Run-I a new Higgs boson-like particle has been found. Furthermore, no new particle has been found up to 600-700\,GeV for generic searches~\cite{searches}. So, in fig.~\ref{gap} we have the picture of the Electroweak Symmetry Breaking Sector (EWSBS) until now: the three would-be Goldstone bosons $\omega^i$ ($i=1,\,2,\,3$) and the recently discovered light Higgs-like scalar $h$.

The presence of the mass gap naturally suggests that this Higgs-like particle could be understood as an additional Goldstone boson (composite state) resulting from a strongly interacting EWSBS dynamics. Different models, like the MCHM (Minimal Composite Higgs Model) based on the SO(5)/SO(4) coset, or the dilaton models (based on the spontaneous breaking of scale invariance symmetry), develop this kind of idea.


To simplify the amplitudes enormously we use the Equivalence Theorem~\cite{ET},
\begin{equation}\label{eq_ET}
T(\omega^a\omega^b \rightarrow \omega^c \omega^d) = T(W_L^a W_L^b\rightarrow W_L^c W_L^d)
+ O\left(\frac{M_W}{\sqrt{s}}\right),
\end{equation}
that restricts us to energies of order $\sqrt{s}\sim 1\,{\rm TeV}$ (the natural upper limit of validity being $4\pi v\sim 3$ TeV). Furthermore, as $m_W \sim m_H \ll \sqrt{s}$, as required by the Equivalence Theorem, we set $m_H = m_W = 0$.

First we analyse, following~\cite{OURS}, the elastic and inelastic scattering of would-be Goldstone bosons (WBGSs). That is, the channels $\omega\omega\rightarrow\omega\omega$, $\omega\omega\rightarrow hh$ and $hh\rightarrow hh$ (this last one, 
though hardly observable at the LHC, is needed to satisfy unitarity). The isospin basis has been used for the three WBGSs $\omega^a$ ($a=1,2,3$). The transformation to the charge basis is $\omega^\pm = (\omega^1\mp i\omega^2)/\sqrt{2}$, $\omega^0=\omega^3$. We employed the non-linear effective Lagrangian with the three WBGSs and a Higgs-like light scalar $h$, with the $W_L W_L$ scattering described by the relevant Lagrangian, invariant under parity, custiodial isospin, and up to dimension 8 in fields and derivatives,
\begin{eqnarray} \label{Lagrangian}
{\cal L} & = & \left( 1\! +\! 2 a \frac{h}{v} +b \frac{h^2}{v^2}\right)
\frac{\partial_\mu \omega^a \partial^\mu \omega^b}{2}
\left( \delta^{ab}\!+\! \frac{\omega^a\omega^b}{v^2} \right)   
\nonumber  \\ 
 & + &  \frac{4 a_4}{v^4}\left( \partial_\mu \omega^a\partial_\nu \omega^a \right)^2 
+ \frac{4 a_5}{v^4}\left( \partial_\mu \omega^a \partial^\mu \omega^a\right)^2  
 \nonumber  \\
 & + & \frac{2 d}{v^4} \partial_\mu  h \partial^\mu h\partial_\nu \omega^a  \partial^\nu\omega^a
+\frac{2 e}{v^4}\left( \partial_\mu h \partial^\mu \omega^a\right)^2
  \nonumber\\
 & + & \frac{1}{2}\partial_\mu h \partial^\mu h +\frac{g}{v^4} (\partial_\mu h \partial^\mu h)^2 \,,
\end{eqnarray}
where the latin lowercase $a\dots g$ coefficients represent BSM couplings.


We now gauge the SM group to extend the Lagrangian density to couple the transverse gauge bosons and the photon following~\cite{2gamma}, so we can describe the photoproduction processes the $\gamma\gamma\rightarrow \omega^+ \omega^-$ and $\gamma\gamma\rightarrow\omega^0\omega^0$. The relevant terms of the LO Lagrangian are
\begin{eqnarray}
{\cal L}_2 &=& -\frac{1}{2 g^2} {\rm Tr}(\hat{W}_{\mu\nu}\hat{W}^{\mu\nu}) -\frac{1}{2 g^{'2}} {\rm Tr} (\hat{B}_{\mu\nu} \hat{B}^{\mu\nu}) \nonumber\\
 &+& \frac{v^2}{4}\left[%
  1 + 2a \frac{h}{v} + b \frac{h^2}{v^2}\right] {\rm Tr} (D^\mu U^\dagger D_\mu U ) \nonumber\\
 &+& \frac{1}{2} \partial^\mu h \, \partial_\mu h
 + \dots\, ,
\label{eq.L2}
\end{eqnarray}
and at NLO,
\begin{eqnarray}
{\cal L}_4 &=&
  a_1 {\rm Tr}(U \hat{B}_{\mu\nu} U^\dagger \hat{W}^{\mu\nu})
  + i a_2 {\rm Tr} (U \hat{B}_{\mu\nu} U^\dagger [V^\mu, V^\nu ]) \nonumber\\
  &-& i a_3  {\rm Tr} (\hat{W}_{\mu\nu}[V^\mu, V^\nu])
 -\frac{c_{\gamma}}{2}\frac{h}{v}\tilde{e}^2 A_{\mu\nu} A^{\mu\nu}\, +\, ... \,,
\label{eq.L4}
\end{eqnarray}
with standard notation
\begin{eqnarray}
D_\mu U &=& \partial_\mu U + i\hat{W}_\mu U - i U\hat{B}_\mu \\
V_\mu &=& (D_\mu U) U^\dagger \\
\hat{W}_{\mu\nu} &=& \partial_\mu \hat{W}_\nu - \partial_\nu \hat{W}_\mu + i  [\hat{W}_\mu,\hat{W}_\nu ]\\
\hat{B}_{\mu\nu}  &=& \partial_\mu \hat{B}_\nu -\partial_\nu \hat{B}_\mu\\
\hat{W}_\mu &=& g W_\mu^a  \tau^a/2 \\
\hat{B}_\mu &=& g'\, B_\mu \tau^3/2 .
\end{eqnarray}
Here, $\tilde{e}$ is the electric charge. To reduce this Lagrangian to a form analogous to Eq.~(\ref{Lagrangian}) in terms of the low-energy quanta directly we have used two approaches corresponding to different representations of the matrix field $U$ describing the WBGBs $\omega^\pm$ and $\omega^0$. The exponential one,
\begin{equation}
 U(x) = \exp\left(i\frac{\tau^a\omega^a(x)}{v}\right),\; a=1,\,2,\,3
\end{equation}
and the more efficient spherical parameterization
\begin{equation}
 U(x) = \sqrt{1-\frac{1}{v^2}\omega^a\omega^a} + i\frac{\omega^a\tau^a}{v},
\end{equation}
where in both cases $\tau^a$ are the Pauli matrices. As  expected, in spite of the different Feynman rules (detailed in ref.~\cite{2gamma}), the resulting on-shell matrix elements for the reactions involving two photons are equal. For the sake of simplicity, we will reproduce here only the simplest effective Lagrangian (without the NLO terms), corresponding to the spherical parametrization, 
\begin{eqnarray}
{\cal L}_2 & = & \frac{1}{2}\partial_\mu h\partial^\mu h 
            +  \frac{\mathcal{F}}{2}\left(2\partial_\mu\omega^+\partial^\mu\omega^- + \partial_\mu\omega^0\partial^\mu\omega^0\right)\nonumber\\
           & + & \frac{\mathcal{F}}{2v^2}\left(\partial_\mu\omega^+\omega^- + \omega^+\partial_\mu\omega^- + \omega^0\partial_\mu\omega^0\right)^2\nonumber\\
           & + & i\tilde{e}\mathcal{F}A^\mu\left(\partial_\mu\omega^+\omega^- - \omega^+\partial_\mu\omega^-\right)\nonumber\\
           & + & \tilde{e}^2\mathcal{F}A_\mu A^\mu\omega^+\omega^- 
\end{eqnarray}
where
\begin{equation}
\mathcal{F} = 1 + 2a\frac{h}{v}+b\frac{h^2}{v^2} \,.
\end{equation}


Many different models can be studied at \emph{low-energy} (i.e., at $\sqrt{s}\sim 1$~TeV) with these effective Lagrangians, by taking different values of the parameters. Particular cases are
\begin{itemize}
  \item $a^2 = b = 1$, SM
  \item $a^2 = b = 0$, Higgsless ECL~\cite{Appelquist:1980vg,Longhitano:1980iz}
  \item $a^2 = 1-\frac{v^2}{f^2}$, $b=1-\frac{2v^2}{f^2}$, $SO(5)/SO(4)$  MCHM~\cite{ref_MCHM}
  \item $a^2 = b = \frac{v^2}{\hat{f}^2}$, Dilaton~\cite{ref_Dilaton}
\end{itemize}
Notice that the $f$ parameter is a new energy scale whose precise meaning depends on the model considered. Besides, for the SM case, the exposed Lagrangians should be extended to include $M_W$, $M_Z$ and $M_H$, since there is a cancellation which leads to a weakly interacting EWSBS, for which the scattering matrix elements would vanish if we neglected these masses. As noted in ref.~\cite{OURS}, unless $\left|a^2-b\right|\gg 0.12$ or $10\pi^2\left|e\right|\gg 0.12$ the Higgs mass $m_h$ should also be kept.


There being no evidence of double Higgs production, there is no direct bound over the $b$ parameter ($\omega\omega h h$ vertex). The strongest one~\cite{const_CMS_ATLAS} is over the $a$ parameter (the $\omega\omega h$ interaction). At a level of $2\sigma$ (95\%),
\begin{itemize}
  \item CMS\dotfill $a\in [0.88,\,1.15]$
  \item ATLAS\dotfill $a\in [0.96,\,1.34]$
\end{itemize}
In addition to this bound, the presence of a mass gap up to 600-700\,GeV~\cite{searches} sets constraints over any set of parameters predicting a new resonance below this value.

Although only two candidates for $\gamma\gamma$ scattering events have been detected until now by CMS~\cite{CMS_two_photon}, the CMS and ATLAS projects are expected to improve this at Run--II~\cite{ref_prosp_two_photon}, specially with the dedicated forward detectors CMS-TOTEM and ATLAS-AFP.

We now examine the resulting amplitudes, starting by the photoproduction one.
It is remarkable how simple the resulting one-loop amplitude $\mathcal{M}$ for the $\gamma\gamma\to \omega\omega$ reaction is, when compared with the complexity of the intermediate computations,
\begin{eqnarray}
\mathcal{M} &=& i\tilde{e}^2 (\epsilon_1^\mu \epsilon_2^\nu T_{\mu\nu}^{(1)}) A(s,t,u) \nonumber\\
            &+& i\tilde{e}^2 (\epsilon_1^\mu \epsilon_2^\nu T_{\mu\nu}^{(2)}) B(s,t,u) \\
T_{\mu\nu}^{(1)} &=& \frac{s}{2} (\epsilon_1\epsilon_2) - (\epsilon_1 k_2) (\epsilon_2 k_1) \\
T_{\mu\nu}^{(2)} &=& 2s(\epsilon_1\Delta)(\epsilon_2\Delta) - (t-u)^2(\epsilon_1\epsilon_2) \nonumber\\
  && -2(t-u)[(\epsilon_1\Delta)(\epsilon_2 k_1) - (\epsilon_1 k_2)(\epsilon_2\Delta)]\\
\Delta^\mu &=& p_1^\mu - p_2^\mu,
\end{eqnarray}
where, shortening the notation so that  $X(jk)$ stands for $X(\gamma\gamma\rightarrow jk)$,
\begin{eqnarray}
M(zz)_{\rm LO} &=& 0 \\
A(zz)_{\rm NLO} &=& \frac{2ac_\gamma^r}{v^2} + \frac{(a^2-1)}{4\pi^2 v^2} \\
B(zz)_{\rm NLO} &=& 0 \\
A(\omega^+\omega^-)_{\rm LO} &=& 2s B(\omega^+\omega^-)_{\rm LO} \nonumber\\
                             &=& -\frac{1}{t} - \frac{1}{\mu} \\
A(\omega^+\omega^-)_{\rm NLO} &=& \frac{8(a_1^r - a_2^r + a_3^r)}{v^2} + \frac{2ac_\gamma^r}{v^2} \nonumber\\
                              &+& \frac{(a^2-1)}{8\pi^2 v^2} \\
B(\omega^+\omega^-)_{\rm NLO} &=& 0,
\end{eqnarray}

Next we turn at length to the $\omega\omega$ scattering. For this the Inverse Amplitude Method (IAM) has been used in order to unitarize the matrix element, in a regime (strong interactions) where the perturbative approach would break down. These \emph{unitarization methods} have been largely and successfully used for the strongly--interacting QCD case in hadron physics, where there are some results very close to those exposed here~\cite{Khemchandani:2011et}, although at much lower energies.

In order to use the unitarization procedures, we have projected the amplitudes over definite orbital angular momentum (the WBGBs carry zero spin),
\begin{equation}\label{Jprojection}
A_{IJ}(s)=\frac{1}{64\,\pi}\int_{-1}^1\,d(\cos\theta)\,P_J(\cos\theta)\,A_I(s,t,u),
\end{equation}
and taken the chiral expansion
\begin{equation}
A_{IJ}(s)=A^{(0)}_{IJ}(s)+A^{(1)}_{IJ}(s)+\dots ,
\end{equation}
where $A^{(0)}_{IJ}\sim O(p^2)$ and $A^{(1)}\sim O(p^4)$. In more detail,
{\small\begin{eqnarray}\label{expandpartialwave}
   A^{(0)}_{IJ}(s) & = & K s   \nonumber\\
   A^{(1)}_{IJ}(s) & = & s^2\left( B(\mu)+D\log\frac{s}{\mu^2}+E\log\frac{-s}{\mu^2}\right) \,.
\end{eqnarray}}
\begin{figure}[t]
\null\hfill\includegraphics[width=.45\textwidth]{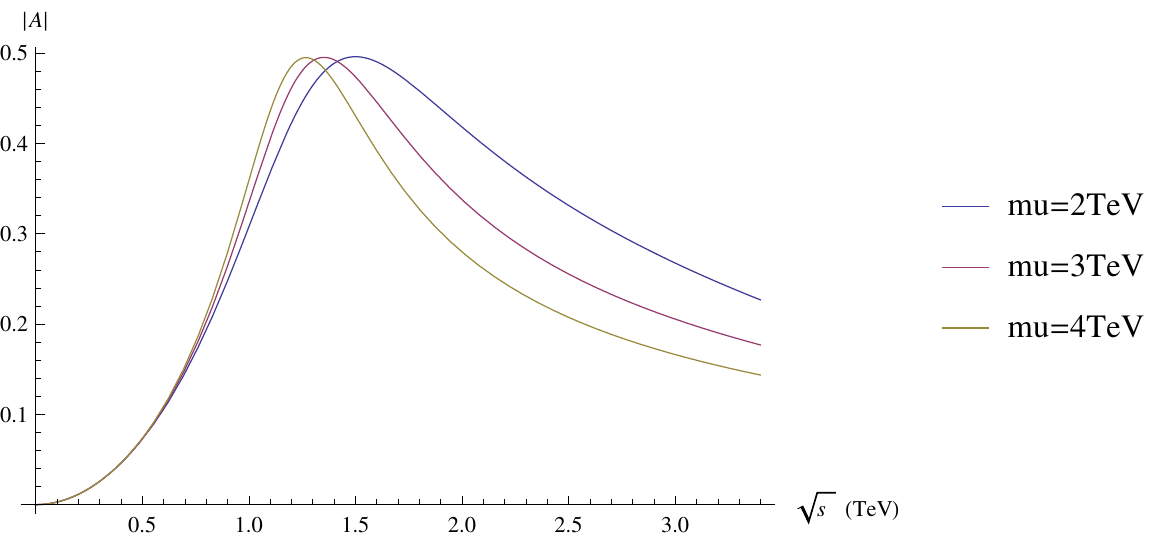}\hfill\null
\caption{Absolute value of the total matrix element of the $\omega\omega\rightarrow hh$ process $\lvert A\rvert$ as function of $\mu$. With $a=1$, $b=2$ and all the other BSM parameters taken to vanish at $\mu$.
}\label{depend_mu}
\end{figure}%
Here, $\mu$ is the renormalization scale which we will take at the cutoff  of the effective theory, $\mu=3\,{\rm TeV}$. Nevertheless, the dependence on this renormalization scale for fixed values of the coupling constants is shown in fig.~\ref{depend_mu} for the elastic $\omega\omega\to \omega\omega$ scattering (of course, if we employ the calculated running values of the parameters from their renormalization group equation the curve remains fixed).

The constants $K$, $D$ and $E$ and the function $B(\mu)$ in Eq.~(\ref{expandpartialwave}) are channel-dependent. For the $\omega\omega\rightarrow\omega\omega$ scattering, and the scalar-isoscalar channel ($IJ=00$),
\begin{eqnarray}{}\label{partial00}
 K_{00} & = & \frac{1}{16 \pi v^2} (1-a^2) \nonumber\\
 B_{00}(\mu) & = &\frac{ 1}{9216 \pi^3 v^4}
  [101(1-a^2)^2 + 68(a^2-b)^2 \nonumber\\
  &+& 768 (7 a_4(\mu) + 11 a_5(\mu)) \pi^2] \nonumber\\
    D_{00} & = & -\frac{1}{4608\pi^3v^4}
  [7(1-a^2)^2 + 3(a^2-b)^2]\nonumber\\
 E_{00} & = & -\frac{1}{1024\pi^3v^4} [4(1-a^2)^2  + 3(a^2-b)^2]\ .
\end{eqnarray}
For the vector isovector ($IJ=11$) one,
\begin{eqnarray}{}\label{partial11}
   K_{11} & = & \frac{1}{96 \pi v^2} (1-a^2) \nonumber\\
   B_{11}(\mu) & = & \frac{1}{110592\pi^3 v^4}
  [8(1-a^2)^2 - 75(a^2-b)^2 \nonumber\\
   &+& 4608 (a_4(\mu) - 2 a_5(\mu)) \pi^2 ] \nonumber\\
   D_{11} & = & \frac{1}{9216\pi^3v^4}
  [(1-a^2)^2 + 3(a^2-b)^2] \nonumber\\
  E_{11} & = & -\frac{1}{9216\pi^3v^4}  (1-a^2)^2\ ;
\end{eqnarray}
it is interesting to note that if we decouple the $hh$ channel by setting $b=a^2$ in this one, $D+E=0$ accidentally. This reflects the known fact from the old electroweak Lagrangian that the counterterm combination appearing in this channel is by itself renormalization group invariant, as can be seen by substituting in Eq.~(\ref{expandpartialwave}) and noting the disappearance of $\mu$ over the physical values of $s$ (right cut). This feature does not remain in the coupled-channel case where $b\neq a^2$, even as isospin-1 cannot couple to $hh$; the effect comes from $hh$ exchange in the $t$-channel.
For the scalar isotensor ($IJ=20$) one,
\begin{eqnarray}{}\label{partial20}
   K_{20} & = & -\frac{1}{32 \pi v^2} (1-a^2) \nonumber\\
   B_{20}(\mu) & = & \frac{1}{18432 \pi^3 v^4}
 [91(1-a^2)^2 + 28(a^2-b)^2 \nonumber\\
  &+& 3072 (2 a_4(\mu) + a_5(\mu)) \pi^2 ]\nonumber\\
   D_{20} & = & -\frac{1}{9216\pi^3v^4}
  [11(1-a^2)^2 + 6(a^2-b)^2]\nonumber\\
  E_{20} & = & -\frac{1}{1024\pi^3v^4}(1-a^2)^2  \ 
\end{eqnarray}
and finally, for the tensor isoscalar ($IJ=02$),
\begin{eqnarray}{}\label{partial02}
   K_{02} & = & 0 \nonumber\\
   B_{02}(\mu) & = & \frac{1}{921600 \pi ^3 v^4}
 [320(1-a^2)^2 + 77(a^2-b)^2 \nonumber\\
  &+& 15360(2 a_4(\mu) + a_5(\mu)) \pi^2 ] \nonumber\\
   D_{02} & = & -\frac{1}{46080\pi^3v^4}
  [10(1-a^2)^2 + 3(a^2-b)^2] \nonumber\\
  E_{02} & = & 0\ .
\end{eqnarray}

If $b\neq a^2$ the channels are coupled and we need
$\omega\omega  \rightarrow h h$, whose scalar partial wave $M_{J=0}$ is
\begin{eqnarray}{} \label{Mscalar}
\nonumber   K'_{0} & = & \frac{\sqrt{3}}{32 \pi v^2} (a^2-b) \\
 \nonumber   B'_{0}(\mu) & = & \frac{\sqrt{3}}{16\pi v^4} \left(d(\mu)+\frac{e(\mu)}{3}\right) \\
 \nonumber &+&\frac{\sqrt{3}}{18432\pi^3 v^4}(a^2-b)[72(1-a^2) + (a^2-b)]    \\
 \nonumber   D'_{0} & = & - \frac{\sqrt{3}(a^2-b)^2}{9216\pi^3 v^4}   \\  
  E'_{0} & = &  -\frac{\sqrt{3}(a^2-b)(1-a^2)}{512\pi^3 v^4}                                   
\end{eqnarray}
with the tensor one $M_{2}$,
\begin{eqnarray}{} \label{Mtensor}
\nonumber   K'_{2} & = & 0 \\
 \nonumber   B'_{2}(\mu)& = & \frac{e(\mu)}{160\sqrt{3}\pi v^4}  +\frac{83(a^2-b)^2}{307200\sqrt{3}\pi^3 v^4}    \\
 \nonumber   D'_{2} & = & - \frac{(a^2-b)^2}{7680\sqrt{3}\pi^3 v^4} \\ 
  E'_{2} & = &  0\  .                                
\end{eqnarray}

Finally for the  $h h \rightarrow  h h$ reaction the $T_{0}(s)$ scalar partial-wave amplitude is given by
\begin{eqnarray}{}\label{Tscalar}
\nonumber   K''_{0} & = & 0 \\
 \nonumber   B''_{0}(\mu) & = & \frac{10g(\mu) }{96\pi v^4}   +  \frac{(a^2-b)^2}{96\pi^3 v^4}   \\
 \nonumber   D''_{0} & = & - \frac{(a^2-b)^2}{512\pi^3 v^4}  \\ 
  E''_{0} & = &  -   \frac{3(a^2-b)^2}{1024\pi^3 v^4}   
\end{eqnarray}{}  
and the tensor $T_{2}$, by
\begin{eqnarray} \label{Ttensor}
\nonumber   K''_{2} & = & 0 \\
 \nonumber   B''_{2}(\mu) & = & \frac{g(\mu) }{240\pi v^4}   +  \frac{77(a^2-b)^2}{307200\pi^3 v^4}   \\
 \nonumber   D''_{2} & = &  -\frac{(a^2-b)^2}{5120\pi^3 v^4}  \\ 
  E''_{2} & = &    0 \ .
  \end{eqnarray}

The unitarity relation for the exact reaction matrix $\tilde{T}$ is
\begin{equation}
  \Imag{\tilde{T}} = \tilde{T}\tilde{T}^\dagger,
\end{equation}
The one loop computation does not satisfy this exactly, only perturbatively. However, following the IAM procedure, we reach an expression
\begin{equation}
A(s)\approx A^{\rm IAM}(s) = \frac{\left(A^{(0)}(s)\right)^2}{A^{(0)}(s)-A^{(1)}(s)}
\end{equation}     

\begin{figure}[th]
\null\hfill\includegraphics[width=.45\textwidth]{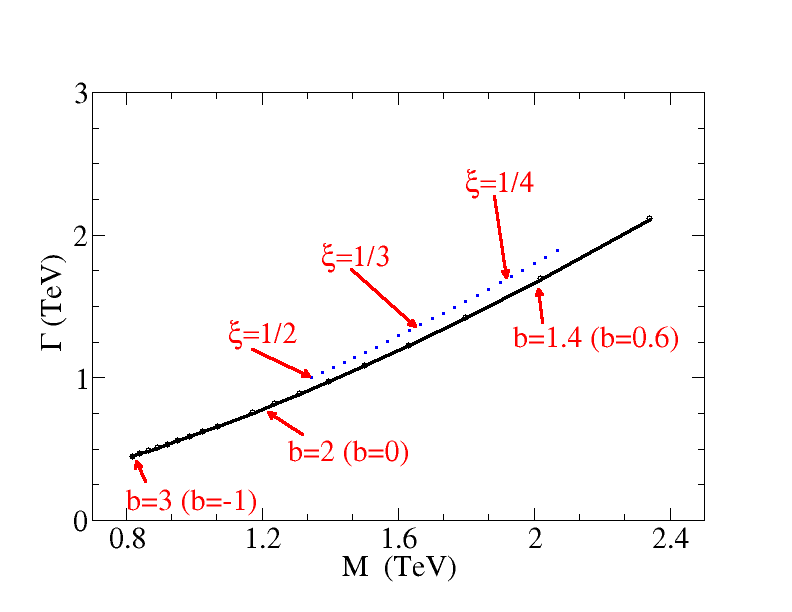}\hfill\null\\
\caption{Dependence of the resonance position on $b$ with $a^2=1$ fixed (upper curve) and for the MCHM (lower curve, blue online), $a=\sqrt{1-\xi}$ and $b=1-2\xi$ ($\xi=v/f$).
}\label{graf_resonance_position}
\end{figure}

\begin{figure}[th]
\includegraphics[width=.4\textwidth]{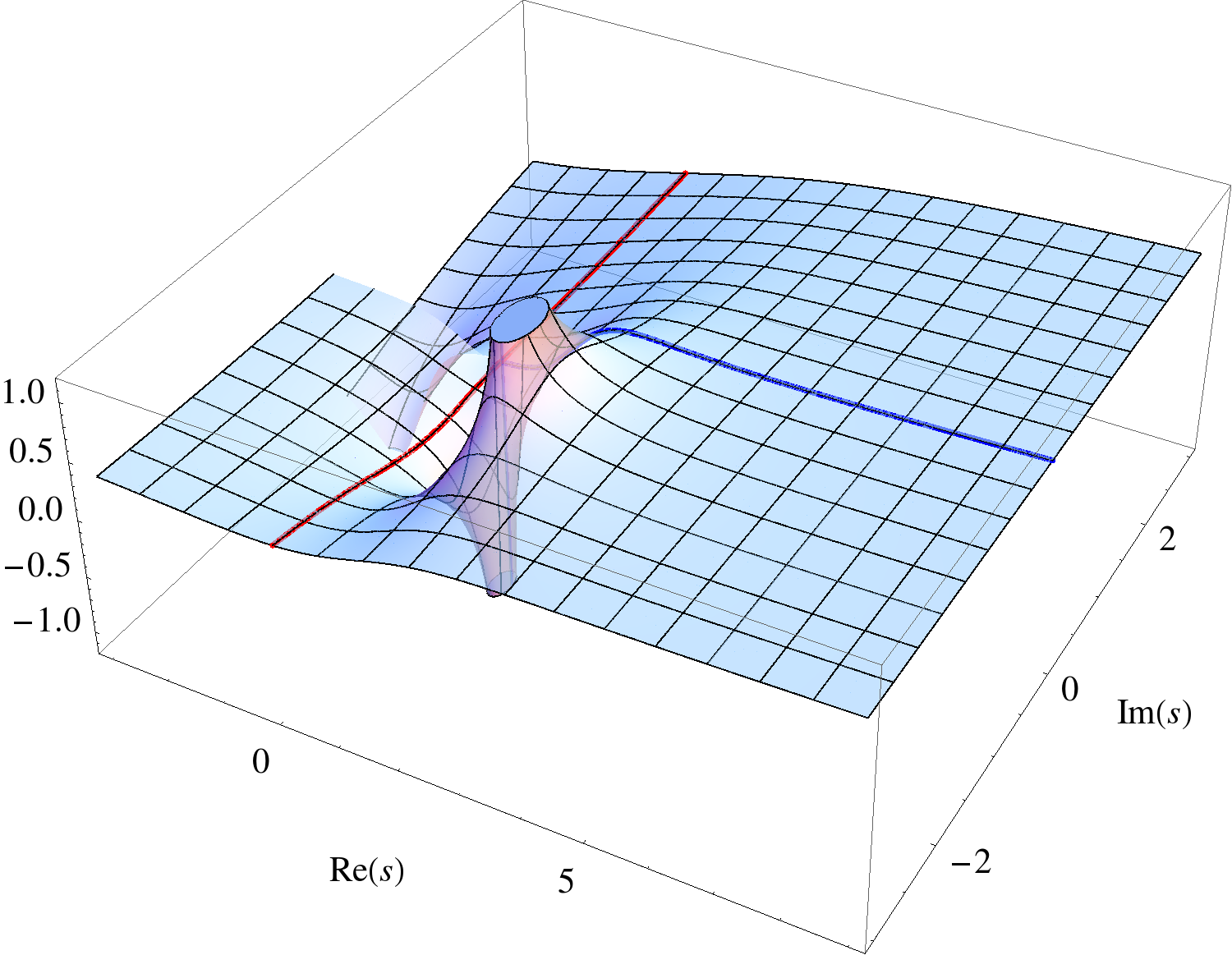}
\caption{Imaginary part of the scattering matrix elements for the $\omega\omega\rightarrow\omega\omega$ channel. With $a=1$, $b=2$, $\mu=3\,{\rm TeV}$ and all the other BSM parameters null.
}\label{pole_second_Riemann_sheet}
\end{figure}

\begin{figure}[th]
\includegraphics[width=.4\textwidth]{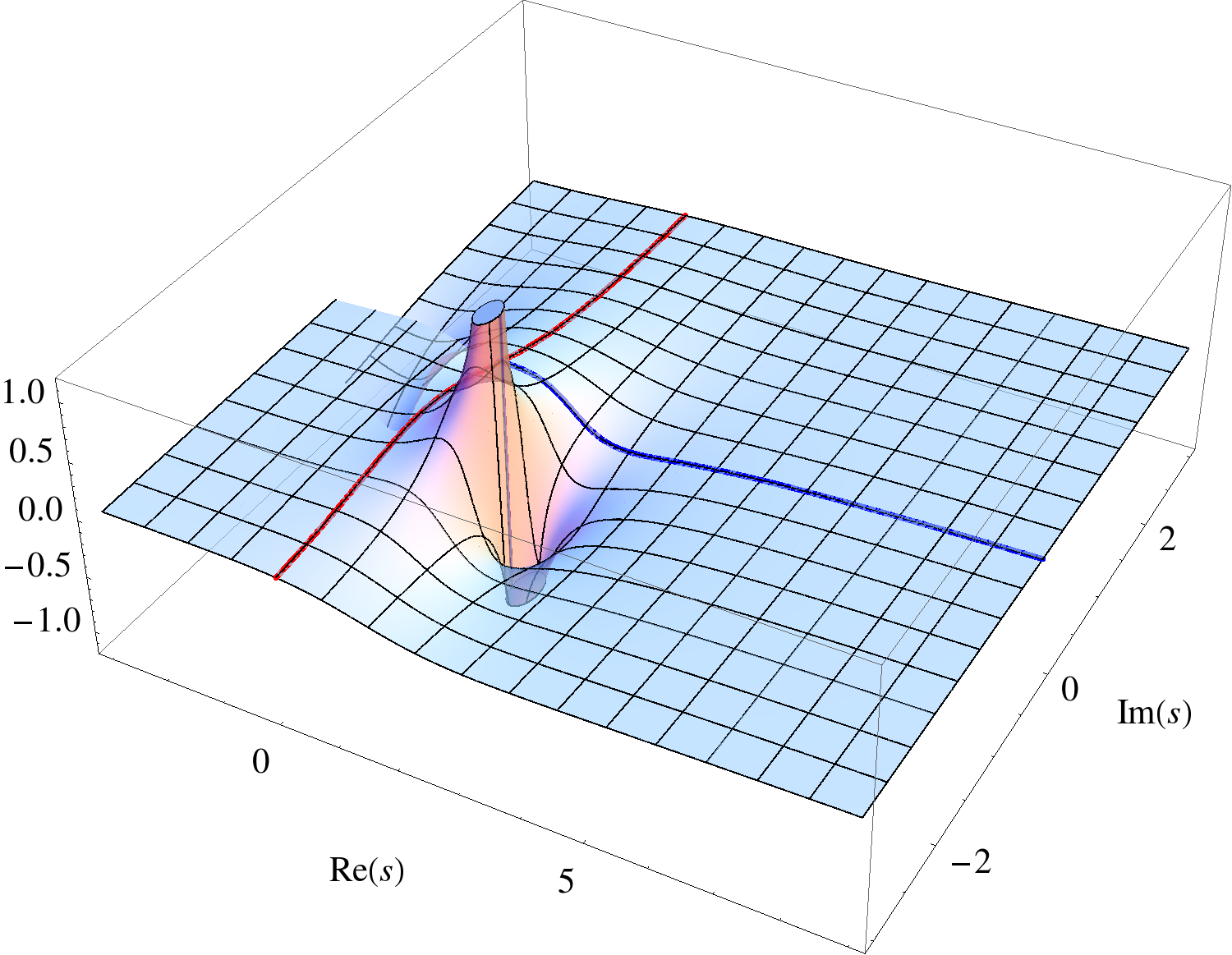}
\caption{Same as in fig.~\ref{pole_second_Riemann_sheet}, but for the $\omega\omega\rightarrow hh$ channel. See that the resonance is in the same position in both cases, as expected.
}\label{pole_second_Riemann_sheet_cross}
\end{figure}

\begin{figure}[th]
\null\hfill\includegraphics[width=.45\textwidth]{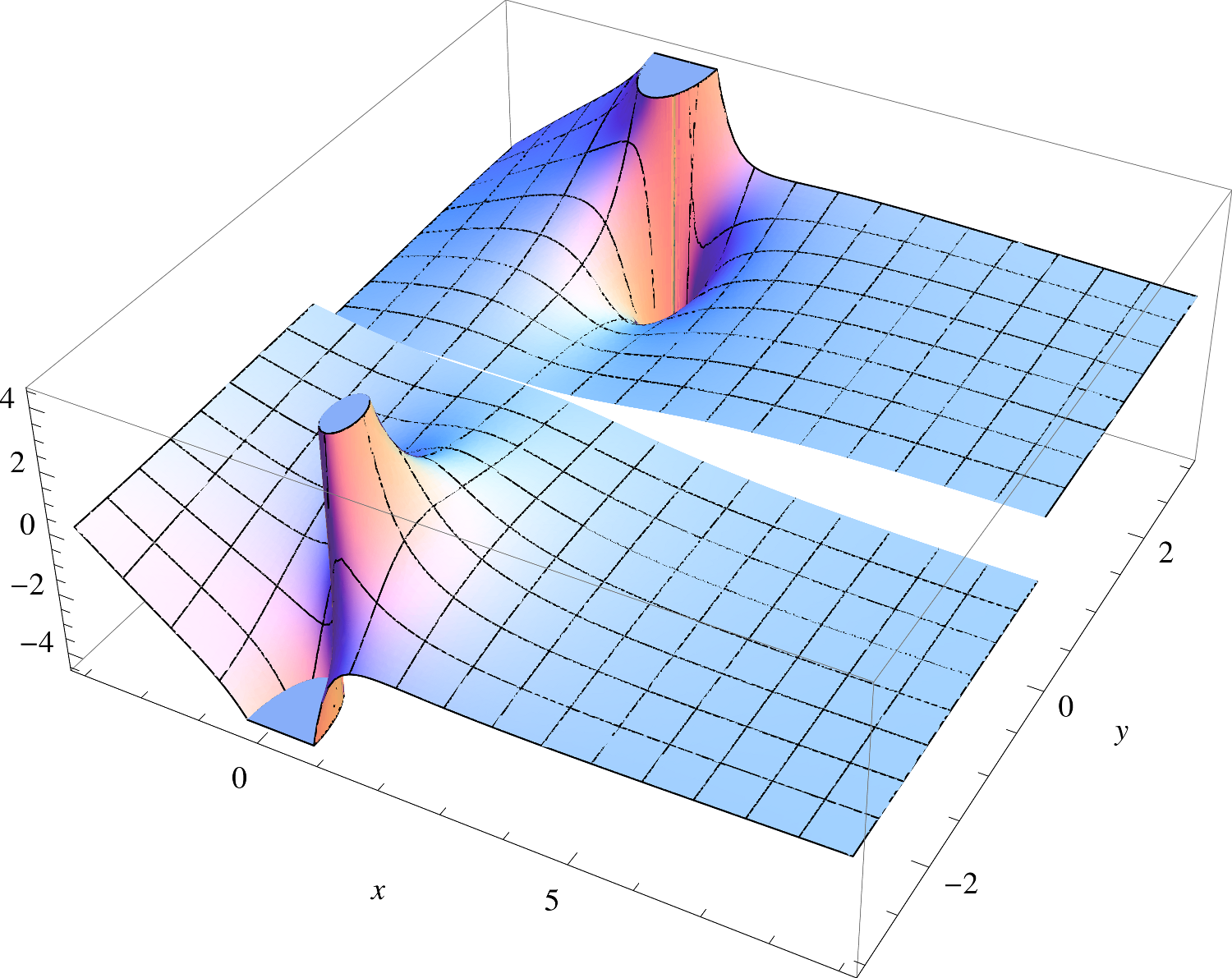}\hfill\null\\
\caption{Pole in the first Riemann sheet for $a=0.9$, $b=1$, $a_4=-0.005$. Isotensor channel ($I=2$, $J=0$).
}\label{graf_Pole_First_Riemann}
\end{figure}

The IAM method, even with $a=1$ and all the other BSM parameters but $b$ vanish, introduces a broad resonance in the TeV scale, whose position is represented in fig.~\ref{graf_resonance_position}. A similar effect has been suggested to occur in the $I=1/2$ resonance oscillating between $\phi N$ and $K^*\Lambda$ around 2 GeV~\cite{Khemchandani:2011et}, due to the strongly interacting QCD in its low energy regime. The appearance of a pole in the second Riemann sheet can be seen in figs.~\ref{pole_second_Riemann_sheet} and~\ref{pole_second_Riemann_sheet_cross}, whereas in fig.~\ref{graf_Pole_First_Riemann} we represent the non--physical appearance of a pole in the first Riemann sheet for a particular set of parameters (which sets the validity limit of the IAM method or of that parameter set).

Two different sets of parameters have been considered. The first one, $a^2=1$, in order for the strong interactions (breaking perturbative unitarity) to come from the 
$\omega\omega\rightarrow hh$ channel (prior to unitarization), a novel effect. As expected, due to the higher order corrections taken into account by the unitarization method, the resonance is also in the same position in the $\omega\omega\rightarrow\omega\omega$ amplitude (figs.~\ref{pole_second_Riemann_sheet} and~\ref{pole_second_Riemann_sheet_cross}). The second one, the Minimal Composite Higgs Model (MCHM, ref.~\cite{ref_MCHM}), $a=\sqrt{1-\xi}$ and $b=1-2\xi$ ($\xi=v/f$). 

Additionally, the authors of~\cite{espriu:2013} studied the reaction to varying $a_4$ and $a_5$. They also studied the regime of parameters for which the unitarization methods give poles in the \emph{first} Riemann sheet, delimitating the validity region of the method.

\begin{figure}[th]
\includegraphics[width=.45\textwidth]{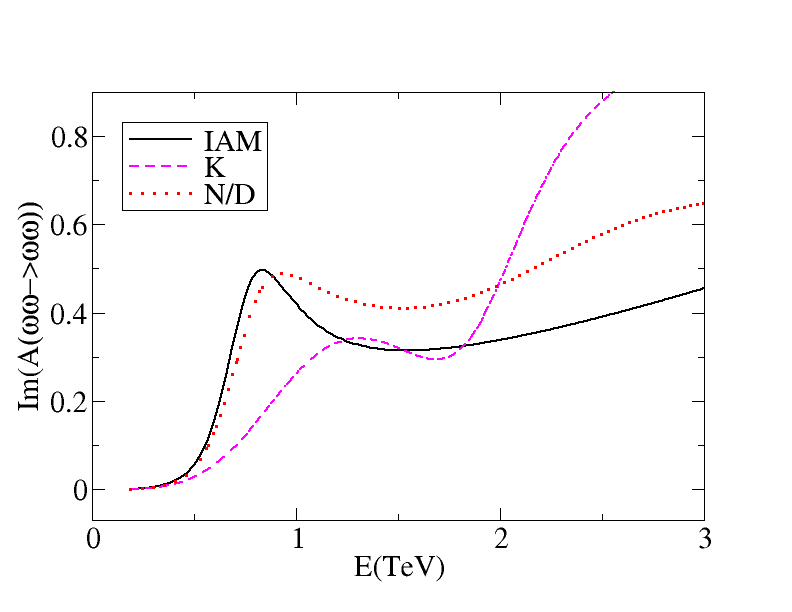}
\caption{Preliminary comparison of two sophisticated ($IAM$, $N/D$) unitarization methods and a very naive one that does not satisfy all proper analytical requirements, the $K$-matrix method. All show the same resonance in the elastic amplitued {\it induced by coupled-channel dynamics} for appropriate $b\neq a^2$.\label{comparison}}
\end{figure}

In figure~\ref{comparison} we show our preliminary computations comparing the IAM with other unitarization methods and that, while specific details may vary,  the prediction of a resonance in the scalar channel for $b\neq 1$ even if $a^2=1$ is robust.


\begin{figure}[t!]
\begin{minipage}{\textwidth}
\includegraphics[width=.45\textwidth]{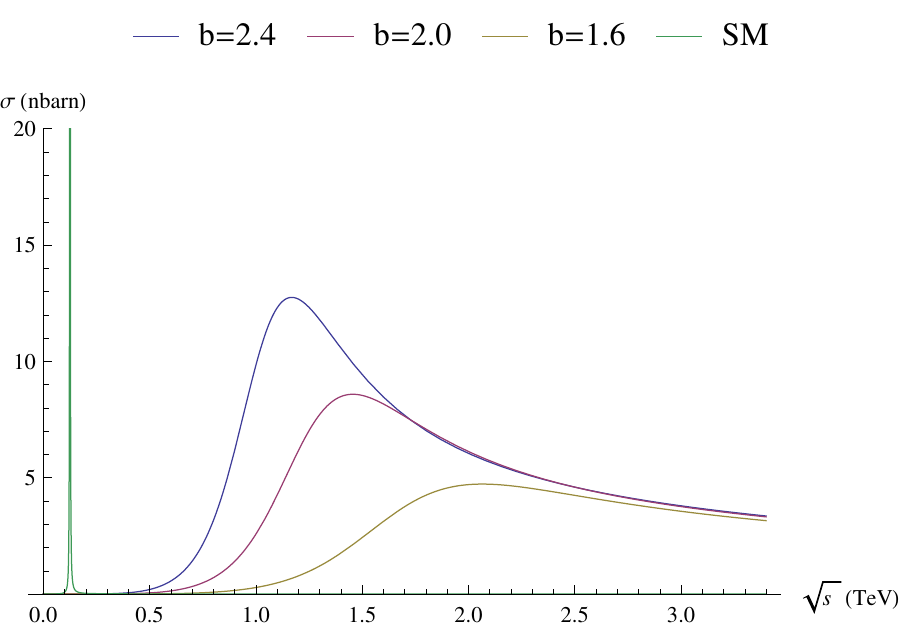}\hfill
\includegraphics[width=.45\textwidth]{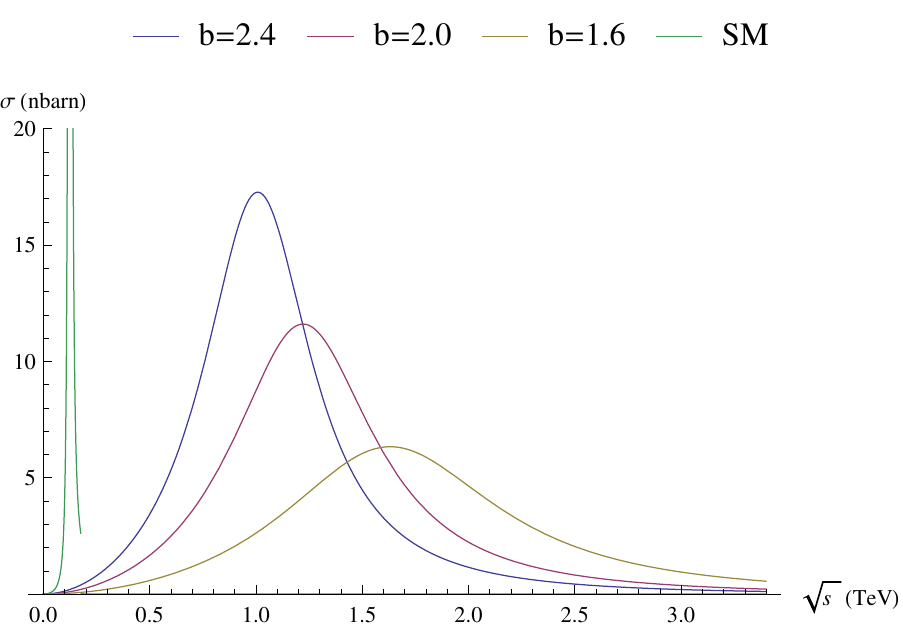}\\
\includegraphics[width=.45\textwidth]{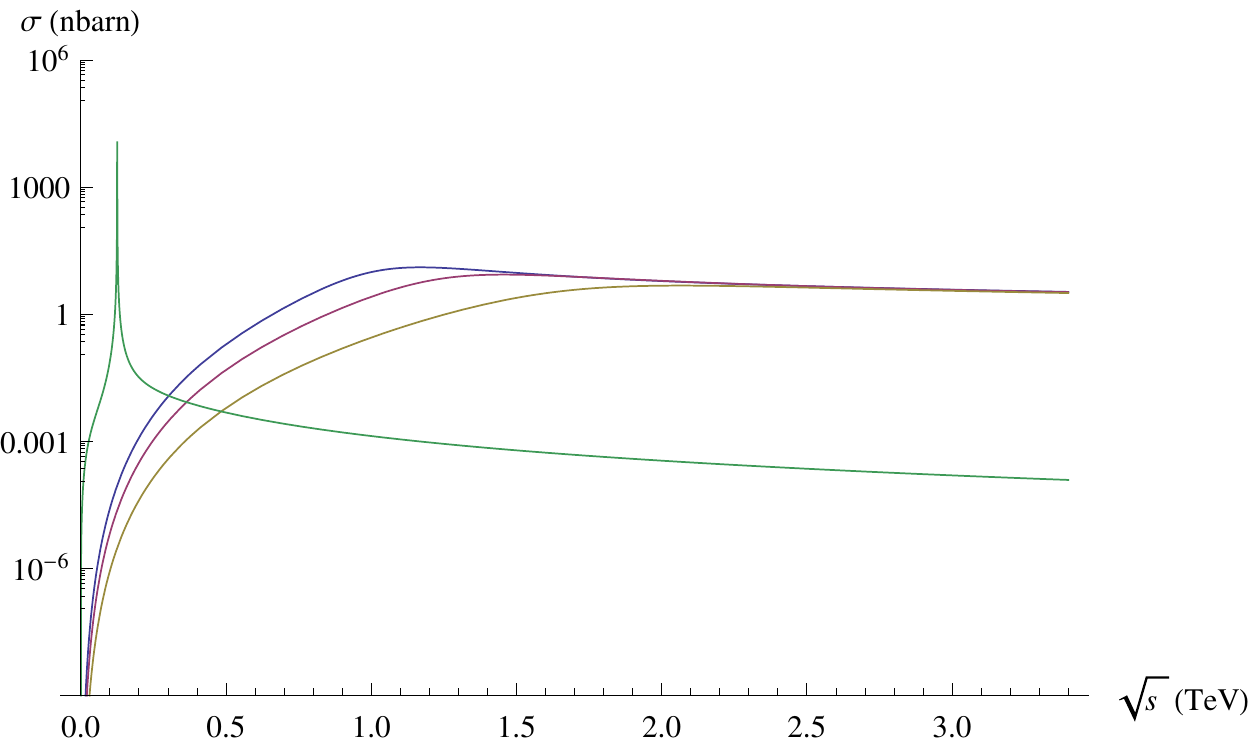}\hfill
\includegraphics[width=.45\textwidth]{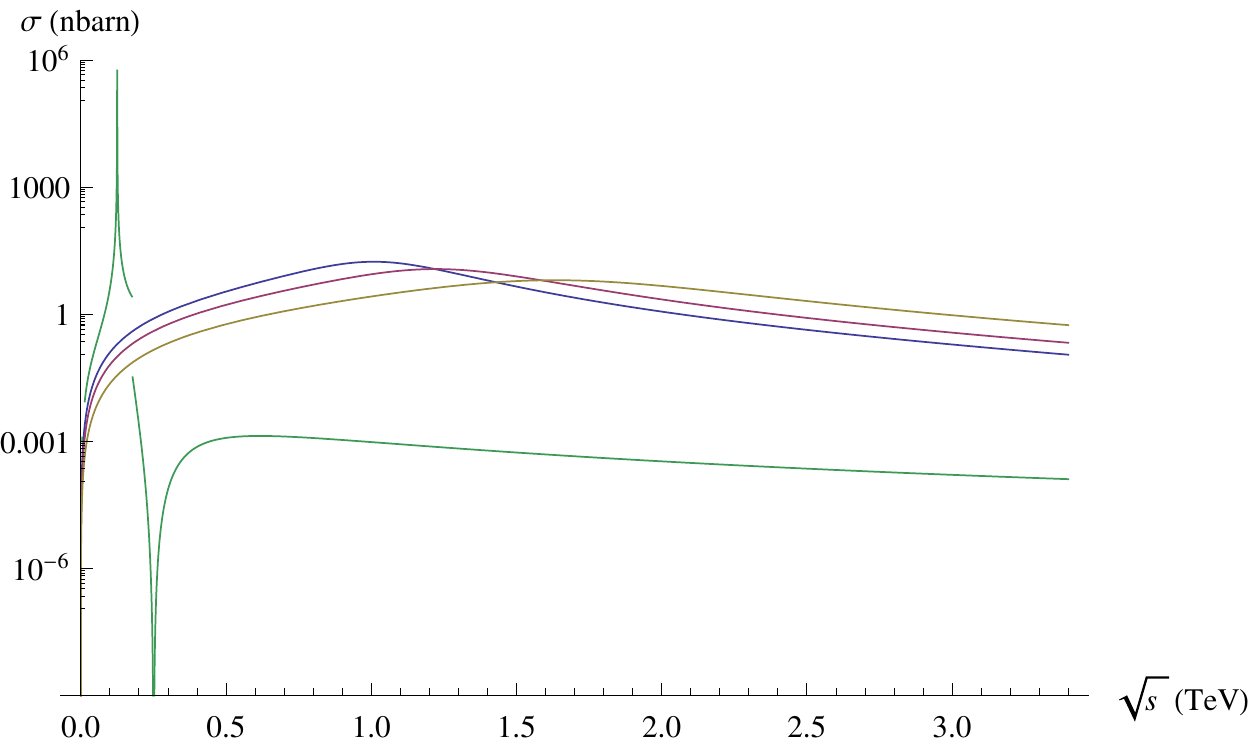}
\end{minipage}\\
\caption{Total cross sections $\omega\omega\rightarrow hh$ (left) and $\omega\omega\rightarrow\omega\omega$ (right) for different values of $b$. With $a=1$, $\mu=3\,{\rm TeV}$ and all the other BSM parameters null. From top to bottom, linear and logarithmic scales for y-axis ($sigma$, in nbarn). Note the enhancement of the cross sections in the strongly interacting scenarios compared with the SM. The BSM computations are not valid for $\sqrt{s}\lesssim 0.6\,{\rm TeV}$ because of the approximations, being one of them the usage of the Equivalence Theorem~\cite{ET}.
}\label{cross_ww}
\end{figure}

\begin{figure}[h!]
\includegraphics[width=.45\textwidth]{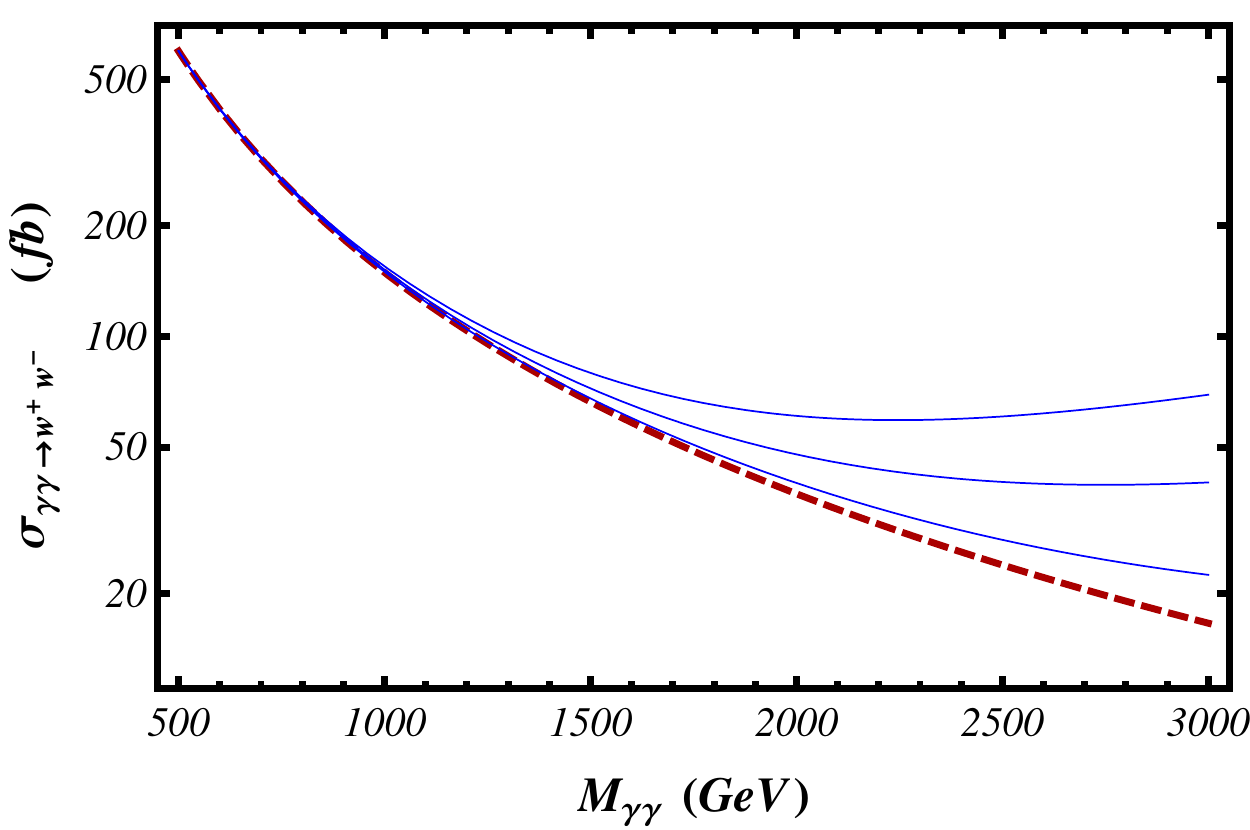}\\
\caption{Total cross section of $\gamma\gamma\rightarrow\omega^+\omega^-$. The red--dashed line correspond to the SM prediction and the solid blue ones our ECLh predictions. From bottom to top, $(a_1 - a_2 + a_3 ) = 2\times 10^{-3},\, 4\times 10^{-3},\, 6\times 10^{-3}$. $a = c_\gamma = 0$.
}\label{cross_gg}
\end{figure}

\begin{figure}[h!]
\null\hfill\includegraphics[width=.45\textwidth]{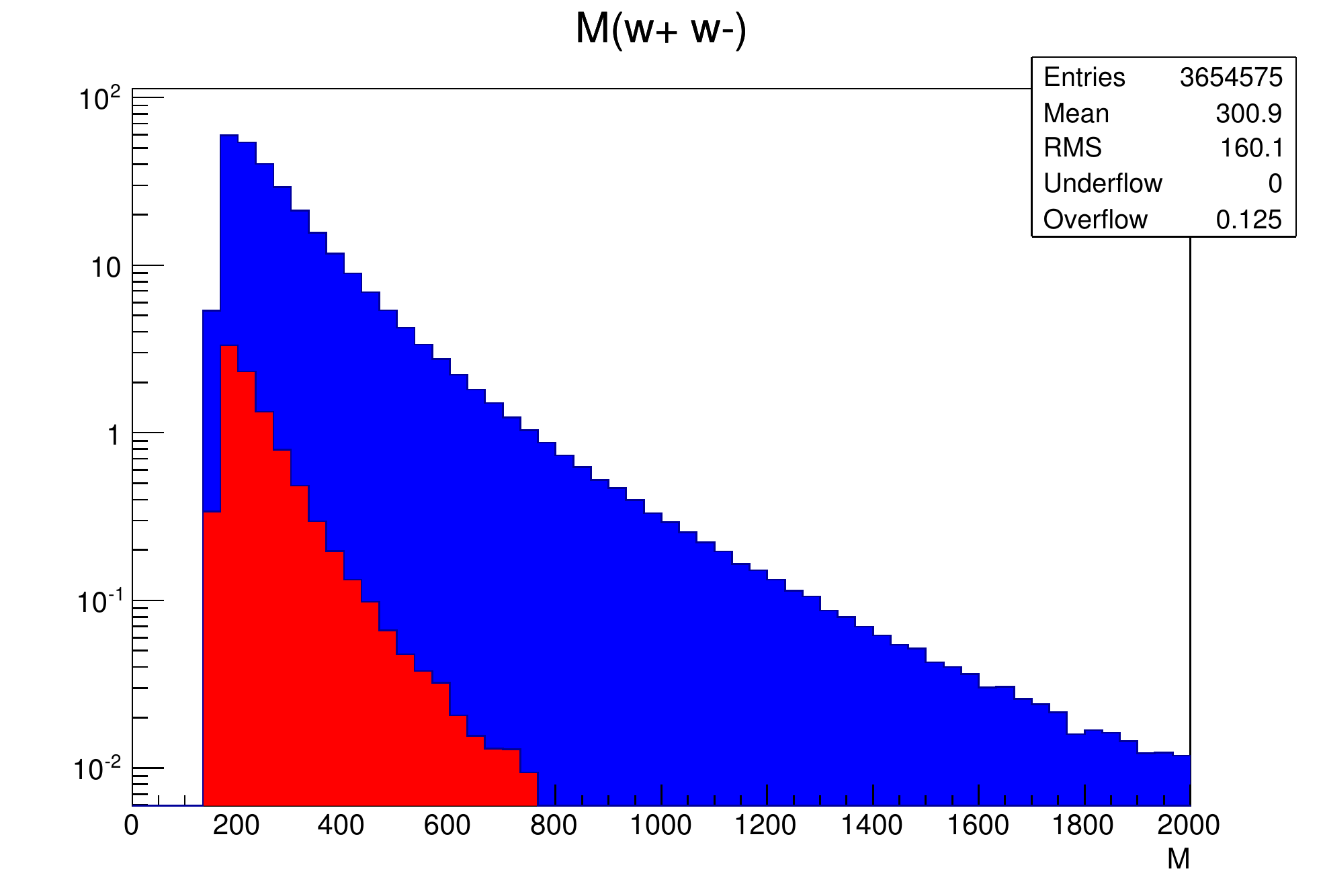}\hfill\null\\
\caption{Monte Carlo computation of the production of W$^+$ W$^-$ (blue) vs. W$_L^+$ W$_L^-$ (red). $\sqrt{s}=\,13\,{\rm TeV}$, $L = 10\,{\rm fb}^{−1}$. x-axis in GeV,
y-axis in events / 33.3 GeV.
}\label{MC}
\end{figure}

The clearest signature of a strongly interacting EWSBS would be an increment on the scattering sections of longitudinal W and Z bosons and photons. Besides the \emph{possible} appearance of QCD--like resonances on the scattering channels. In order to quantify these effects, we have plotted the total cross sections in figs.~\ref{cross_ww} and~\ref{cross_gg}. However, how hard is the experimental challenge? To give an idea, see fig.~\ref{MC}, where we have represented the production of both unpolarized W$^\pm$ and longitudinally polarized W$_L^\pm$. At high energy, experimentally separating the longitudinal ones$\sim \omega\omega$  can be hard as they are a clear minority of the events near SM conditions (strong interactions will bring about deviations from this behavior). 
Perhaps the photon-scattering production, selected by the forward detectors of the LHC detecting an intact proton, or even a future $\gamma\gamma$ collider, could give a clearer experimental signal. 


\begin{theacknowledgments}
The authors thank the usefull suggestions of Mar\'ia Herrero and Juan Jos\'e Sanz-Cillero.
A. Dobado would like to thank useful conversations with D. Espriu and J. R. Pel\'aez for ref.~\cite{Khemchandani:2011et}. 
RLD thanks the hospitality of the NEXT institute and the high energy group at the University of Southampton.  We acknowledge the computer resources, technical expertise, and assistance provided by the BCS and the Tirant supercomputer staff at Valencia. This work is partially supported by the 
CICYT through the project
FPA2011-27853-C02-01, 
by the Spanish Consolider-Ingenio 2010 Programme CPAN (CSD2007-00042) 
and by 
the Spanish MINECO supporting the work of R.L. Delgado under grant BES-2012-056054.
\end{theacknowledgments}


\end{document}